\documentclass[10pt]{article}
\usepackage[cp1251]{inputenc}
\newcommand{\sss}{\scriptstyle}
\usepackage{graphicx}
\usepackage[dvips]{epsfig}
\def\lsim{\  \lower-1.2pt\vbox{\hbox{\rlap{$<$}\lower5pt\vbox{\hbox{$\sim$}}}}\ }
\def\gsim{\  \lower-1.2pt\vbox{\hbox{\rlap{$>$}\lower5pt\vbox{\hbox{$\sim$}}}}\ }

\begin{document}
\title{Thermodynamics of a one-dimensional system of point bosons:\\
 comparison of the traditional approach with a new one}
\author{ {\small Maksim Tomchenko}
\bigskip \\ {\small Bogolyubov Institute for Theoretical Physics, } \\
 {\small 14b,  Metrolohichna Str., Kyiv 03680, Ukraine} \\
 {\small E-mail:mtomchenko@bitp.kiev.ua}}
 \date{\empty}
 \maketitle
 \large
 \sloppy

\begin{abstract}
We compare two approaches to the construction of the thermodynamics
of a one-dimensional periodic system of spinless point bosons:  the
Yang--Yang approach and a new approach proposed by the author. In
the latter, the elementary excitations are introduced so that there
is only one type of excitations (as opposed to Lieb's approach with
two types of excitations: particle-like and hole-like). At the weak
coupling, these are the excitations of the Bogolyubov type. The
equations for the thermodynamic quantities in these approaches are
different, but their solutions coincide (this is shown below and is
the main result). Moreover, the new approach is simpler. An
important point is that the thermodynamic formulae in the new
approach for any values of parameters are formulae for an ensemble
of quasiparticles with the Bose statistics, whereas a formulae in
the traditional Yang--Yang approach have the Fermi-like one-particle
form.
\end{abstract}

 \section{Introduction}
The modern physics of one-dimensional (1D) many-particle systems is
divided, in fact, into two areas: physics of point particles and
physics of nonpoint ones. Moreover, these areas are not fairly
joined, and not all connections between them are clear. Lieb's
approach \cite{lieb1963} involves  particle-like and  hole-like
excitations. The dispersion law of the particle-like excitations
coincides with that for a system of nonpoint bosons
\cite{bog1947,bz1956,fc,brueckner} (though, this is verified only
for a weak coupling and periodic boundary conditions). But none of
the models of a system of nonpoint bosons find the hole-like
excitations (see reviews \cite{rea,mtreview}). This is the first
mismatch. On the basis of the Lieb's approach, Yang and Yang
constructed the thermodynamics of a system of point bosons
\cite{yangs1969}, where the formulae have the Fermi-like form,
though the thermodynamics of a system of nonpoint bosons is
described by the formulae for an ensemble of Bose quasiparticles
\cite{huang}. This is the second mismatch. These two mismatches of
the theories of point and nonpoint particles are usually referred to
(1) particular properties of a point interaction in the 1D case and,
sometimes, (2) the drawbacks of the models for nonpoint particles.
Note that some models of a system of nonpoint particles
\cite{bog1947,brueckner} use a condensate. We recall that a uniform
1D system of nonpoint bosons can possess a quasicondensate close to
the true condensate, if the system is finite, the coupling is very
weak, and the temperature is low \cite{mtcond2016}.

For the fermions, the hole-like excitations are a characteristic
property in the cases of point and nonpoint interactions. The
history and the analysis of models of a 1D system of point fermions
can be found in \cite{gaudinm,takahashi,essler2005,guan2013}.

A new way of introduction of elementary excitations for a system of
point bosons is proposed in the recent work \cite{mt2015}. In this
approach, there is only \textit{one} type of excitations and the
thermodynamic formulae coincide with those for nonpoint bosons.
However, no comparison of the thermodynamics with that in the
traditional Yang--Yang  approach \cite{yangs1969} was made. Below,
we will carry out such comparison and will see that the new
\cite{mt2015} and traditional \cite{yangs1969} approaches are
equivalent. However, the new approach is simpler and removes both
above-mentioned mismatches of the theories of point and nonpoint
bosons. We note that the thermodynamics of 1D systems of point
bosons was developed
\cite{ying2001,taka2007,batchelor2011,guan2015}, but these methods
are based on the Yang--Yang approach \cite{yangs1969}. However, the
approach  \cite{mt2015} is different in essence.

\section{Basic equations and ways of introduction of quasiparticles}
In the present work, we will compare the traditional Yang--Yang
approach \cite{yangs1969} to the thermodynamics of spinless point
bosons with the new one \cite{mt2015}. The Yang--Yang approach was
analyzed in \cite{takahashi} in detail. These approaches differ by
different ways of introduction of elementary excitations. Therefore,
we consider firstly these ways. The system of $N$ point bosons is
described by the Schr\"{o}dinger equation \cite{ll1963}
\begin{equation}
 -\sum\limits_{j}\frac{\partial^{2}}{\partial x_{j}^2}\Psi + 2c\sum\limits_{i<j}
\delta(x_{i}-x_{j})\Psi=E\Psi,  \quad i,j =1,\ldots, N.
     \label{1-1} \end{equation}
Here, we take $c>0$ and use the units with $\hbar =2m=k_{B}=1$. The
theory of penetrable point bosons started by the classical works by
Lieb and Liniger \cite{ll1963} and by Lieb \cite{lieb1963}. In work
\cite{ll1963}, the following equations for quasimomenta $k_{i}$ of a
periodic system of $N$ bosons were found:
\begin{equation}
 (-1)^{N-1}e^{-ik_{j}L}=
 \exp{\left (-2i\sum\limits_{s=1}^{N}\arctan{\frac{k_{s}-k_{j}}{c}}\right )}, \quad  j =1,\ldots, N,
     \label{1-2} \end{equation}
where $L$ is the size of the system. The analysis of work
\cite{ll1963} is based on the equation for the quantity
$k_{j+1}-k_{j}$. It was shown \cite{yangs1969} that Eqs. (\ref{1-2})
yield the equations
\begin{eqnarray}
Lk_{j}=2\pi
I_{j}-2\sum\limits_{l=1}^{N}\arctan{\frac{k_{j}-k_{l}}{c}},   \quad
j =1,\ldots, N.
     \label{1-3} \end{eqnarray}
The ground state of the system corresponds to the quantum numbers
$I_{j}\equiv I^{(0)}_{j}=j-\frac{N+1}{2}$ \cite{gaudinm,gaudin1971}.
For such quantum numbers, Eqs. (\ref{1-3}) are similar to the
equations for the ``Fermi sea,'' and the elementary excitations can
be divided into hole-like and particle-like \cite{lieb1963}.
However, Gaudin noticed \cite{gaudinm,gaudin1971} that, with the
help of the  equality $
\arctan{\alpha}=(\pi/2)sgn(\alpha)-\arctan{(1/\alpha)},$ Eqs.
(\ref{1-3}) can be written in the form
\begin{eqnarray}
Lk_{j}=\left. 2\pi
n_{j}+2\sum\limits_{l=1}^{N}\arctan{\frac{c}{k_{j}-k_{l}}}\right|_{l\neq
j}, \quad j =1,\ldots, N,
     \label{1-4} \end{eqnarray}
where $ n_{j}$ are integers,  and the equality
$I_{j}=n_{j}+j-\frac{N+1}{2}$ holds. Equations (\ref{1-4}) follow
from (\ref{1-3}) at the ordering $k_{1}<k_{2}<\ldots < k_{N}$, which
requires that the inequalities $I_{1}<I_{2}<\ldots < I_{N}$ and
$n_{1}\leq n_{2}\leq \ldots \leq n_{N}$ be satisfied. For the ground
state, all quantum numbers $ n_{j}$ are the same: $ n_{j}=0$.
Therefore, all $N$ equations (\ref{1-4}) are equivalent. This
corresponds to the Bose symmetry. On the basis of Eqs. (\ref{1-4}),
a quasiparticle is defined very simply \cite{mt2015}: The elementary
excitation (quasiparticle) is associated with a change in one of
$n_{j}$ by any integer (as compared with the ground state, for which
$ n_{j}=0$ for all $j$). A change in $ n_{j}$ with the other $j$
means the creation of one more quasiparticle, and so on. In this
case, index $j$ enumerates quasiparticles, and the value of $ n_{j}$
characterizes the $j$-th quasiparticle, by determining its momentum
and energy (see \cite{mt2015}). For such way of introduction of
quasiparticles, it is possible to calculate the statistical sum for
the system with $N=\infty, L = \infty$. As a result,  the same
formula for the total free energy, as for the one-dimensional He II
\cite{huang}, is obtained \cite{mt2015}:
\begin{equation}
F= E_{0} +T\sum\limits_{l\neq 0}\ln{\left (1-e^{-E(l)/T}\right )}.
     \label{1-5} \end{equation}
Here, $E_{0}$ is the ground-state energy of the system, and $E(l)$
means the energy of a quasiparticle (for the free bosons, the same
formula \cite{huang,land5} is true, where $E_{0}=0$ and $E(l)$ is
the energy level of a boson). For the periodic boundary conditions
(BCs), $l$ runs the same values as any $n_{j}$ in system
(\ref{1-4}): $l=\pm 1, \pm 2, \ldots$ ($l=0$ does not enter
(\ref{1-5}), because zero $ n_{j}$ corresponds to the absence of an
excitation). For the zero BCs, formula (\ref{1-5}) is valid as well,
but $l$ runs the values $ 1, 2, 3, \ldots$ \cite{mt2015}. In the
limit $N, L \rightarrow \infty$, $N/L=const,$ the energy levels of
the system with periodic and zero BCs coincide \cite{mt2015}. It
follows from (\ref{1-5}) that the values of thermodynamic parameters
for such systems also  coincide. The assertions made for zero BCs
are valid, if the equations under zero BCs (they are similar to
(\ref{1-4})) have the unique solution; the uniqueness was proved in
\cite{mt2016edinst}. Let us call the method \cite{mt2015} the
$n$-approach,  in view of $n_{j}$ in Eqs. (\ref{1-4}).

We note that the procedure in \cite{mt2015} cannot be repeated on
the basis of Eqs. (\ref{1-3}). This is related to the following. The
ground state corresponds to the collection of successive integer (or
half-integer) numbers $I^{(0)}_{j}$. To obtain an elementary
excitation, one of the numbers $I^{(0)}_{j}$ should be changed by an
integer. But if the changed $I_{l}$ will coincide with one of the
remaining $I^{(0)}_{j}$, we obtain the solution with two identical
$k_{j}$, which is forbidden \cite{ll1963}. In order that the changed
$I_{l}$ will not coincide with one of the remaining $I^{(0)}_{j}$,
we should take out $I_{l}$ outside the bounds of the ``Fermi sea''
of the numbers $I^{(0)}_{j}$. This leads to hole-like and
particle-like excitations of the Lieb's picture. However, the
transition to the system (\ref{1-4}) changes the situation
radically, because the coincidence of several $n_{j}$ is admissible
for (\ref{1-4}): This does not lead to the coincidence of $k_{j}$,
which allows us to introduce quasiparticles with the Bose
statistics. Thus, the key point in the definition of quasiparticles
is the structure of the equations for $k_{j}$.

It is of importance to understand the interconnection between
different approaches: particle-like and hole-like excitations by
Lieb, ``holes'' and $k$'s by Yang and Yang \cite{yangs1969}, and
quasiparticles in the $n$-approach. In the literature, $k$'s by Yang
and Yang are frequently called ``particles.'' It is probably not
quite suitable term, since such ``particle'' is sometimes confused
with Lieb's particle-like excitation. The latter is the excitation
of the whole system, which arises at the transfer of $I_{j}$ from
the ``Fermi sea'' surface outward (in this case, $I_{N}$ increases
by a natural number $s$, or $I_{1}$ decreases by $s$). The hole-like
excitation by Lieb corresponds to the transfer of $I_{j}$  from the
depth of the Fermi sea to the surface (this is equivalent to the
increase in \textit{several} numbers $I_{N-l},
I_{N-l+1},\ldots,I_{N}$ by $1$). The hole- and particle-like
excitations are similar, respectively, to a ``hole'' inside the
Fermi sea and to a hole on the Fermi sea surface.  Yang and Yang
\cite{yangs1969} referred a hole to a change in any $I_{j}$ and did
not use the term ``particle.'' If $I^{(0)}_{l}$  from the collection
$\{I^{(0)}_{j}\}$ changes, then all $k_{j}$ in (\ref{1-3}) change,
and $k_{l}$ changes above all. Therefore, the input value of $k_{l}$
seemed to come out from the distribution of $\{k_{j}\}$. Yang and
Yang associated a ``hole'' with such $k_{l}$. The remaining $k_{j}$
vary slightly and are called $k$'s \cite{yangs1969}. It is worth
noting that, in the Yang--Yang approach, a hole and each $k_{j}$
from $k$'s is a single number, whereas the excitations by Lieb are
collective, since they include changes in all numbers $k_{j}$.

Consider two simple examples. Equations (\ref{1-4}) with $n_{j}=0$
for all $j$ describe the ground state. Let us increase $n_{N}$ by
$1$. According to the $n$-approach, this creates a quasiparticle. In
view of the relation $I_{j}=n_{j}+j-\frac{N+1}{2}$, our action
increases $I_{N}$ by $1$. Within Lieb's approach, this means the
creation of a particle-like excitation. In the Yang--Yang approach,
the ``hole'' $k_{N}$ is created and the remaining $k$'s are slightly
shifted. We now pass from the last state (with $n_{j\leq N-1}=0$,
$n_{N}=1$) to the state with $n_{j\leq N-2}=0$, $n_{N-1}=n_{N}=1$.
In the $n$-approach language, this means the creation of the second
quasiparticle. In the Yang--Yang approach, the second hole
($k_{N-1}$) appears, and $k$'s ($k_{1},\ldots,k_{N-2}$) are shifted.
Within Lieb's approach, a particle-like excitation disappears and a
hole-like excitation, corresponding to the shift of $I_{N-1}$,
$I_{N}$ by $1$, is created. A hole-like excitation corresponds in
the $n$-approach to several quasiparticles with the same (minimal)
momentum.  Lieb's language is not very suitable due to a complicated
connection of quasiparticles with quantum numbers $I_{j}$ and to the
separation of excitations into particle-like and hole-like ones. The
Yang--Yang approach uses also two quantities (holes and $k$'s), and
the visual picture is not quite simple, but such language is
efficient for the construction of the thermodynamics. The
$n$-approach is simpler than the approach by Lieb (due to a single
type of quasiparticles and a simpler connection of quasiparticles
with quantum numbers $n_{j}$) and than the Yang--Yang approach
(since a quasiparticle is considered as a single object, whereas the
approach Yang--Yang considers separately all $k_{j}$ instead of a
single quasiparticle). However, these three approaches are
mathematically equivalent.

Let us pass to the formulae.  Here, we consider a system with
periodic BCs, because the thermodynamics in \cite{yangs1969} was
constructed for a periodic system. First, let us consider the
$n$-approach. It is convenient to pass from formula (\ref{1-5}) to
the equivalent formula \cite{mt2015}
\begin{equation}
F = E_{0} +\frac{TL}{2\pi}\int\limits_{-\infty}^{\infty}\hbox{d}p
\ln{\left (1-e^{\frac{-E(|p|)}{T}}\right )},
     \label{1-6} \end{equation}
where   $E(|p|)$ and $p$ are, respectively, the energy and the
momentum of a quasiparticle. For low $T,$ it is the exact formula
for the free energy of a 1D system of point bosons. As is seen, the
formula is written in the language of quasiparticles. The analogous
formula is known for a system of nonpoint bosons \cite{huang,land5}.
In \cite{land5}, this formula was deduced from properties of the
ensemble of quasiparticles. In the same way, we can obtain formula
(\ref{1-6}) for a system of point bosons, if the quasiparticles are
defined in the $n$-approach (Lieb's quasiparticles are not
characterized by Bose symmetry; therefore, formula (\ref{1-6}) is
wrong for them). We note that formula (\ref{1-6}) was obtained in
\cite{mt2015} in a different way, by the direct summation of the
partition function.

The thermodynamic relation \cite{land5}
\begin{equation}
dF= -SdT -PdV+\mu dN
     \label{1-7} \end{equation}
yields the formula for the total entropy
\begin{equation}
S=-\frac{\partial F}{\partial T}|_{N,V=const}.
     \label{1-8} \end{equation}
Formula (\ref{1-6}) and the dispersion law $E(|p|)$ completely
assign the thermodynamics of the system. Importantly, that formula
(\ref{1-6}) for all values of $T$ and $n$ ($n=N/L$) is a formula for
the gas of noninteracting quasiparticles with Bose statistics. In
this case, the approach \cite{mt2015} involves only one type of
quasiparticles. For all $\gamma = c/n,$ their dispersion law
coincides with the dispersion law of particle-like excitations by
Lieb.

Yang and Yang have obtained the following thermodynamic formulae
\cite{yangs1969}:
\begin{equation}
F = N\mu -PL, \quad P = \frac{T}{2\pi
}\int\limits_{-\infty}^{\infty}dk  \ln{\left
(1+e^{\frac{-\epsilon(k)}{T}}\right )},
     \label{1-9} \end{equation}
\begin{equation}
\epsilon(k) = -\mu +k^{2}-\frac{Tc}{\pi }\int\limits_{-\infty}^{\infty}dq
\frac{\ln{\left (1+e^{\frac{-\epsilon(q)}{T}}\right )}}{c^{2}+(k-q)^{2}},
     \label{1-11} \end{equation}
 \begin{equation}
dP= (S/L)dT +(N/L)d\mu,
     \label{1-12} \end{equation}
\begin{equation}
n\equiv\frac{N}{L}=\frac{\partial P}{\partial \mu}|_{T=const},
     \label{1-10} \end{equation}
\begin{equation}
S=L\frac{\partial P}{\partial T}|_{\mu=const}.
     \label{1-13} \end{equation}
These formulae have the Fermi-like form and are deduced on the basis
of the ideology of the Fermi sea of numbers $I^{(0)}_{j}$.

To calculate the thermodynamic quantities within the $n$-approach
\cite{mt2015}, it is necessary to find $E(p)$ from a linear integral
equation (see Eqs. (2.18)--(2.20) in \cite{lieb1963} or  (43), (44),
 (55) in \cite{mt2015}) and then to determine the free energy $F$
from (\ref{1-6}). In the Yang--Yang approach, the problem is more
complicated: We need to determine $\epsilon(k)$ as a function of
$c,T,\mu$ from the nonlinear integral equation (\ref{1-11}); to find
$\mu(n,T,c)$ from (\ref{1-10}); and then to determine $F$ and the
entropy $S$ from (\ref{1-9}), (\ref{1-13}).

We note that formulae (\ref{1-6}) and (\ref{1-9})--(\ref{1-13}) are
true for $N, L \rightarrow \infty$, $N/L=const$. In this case,
formulae (\ref{1-9})--(\ref{1-13}) are valid for any $T$, but
formula (\ref{1-6}) holds only at low $T$, for which the interaction
of quasiparticles can be neglected \cite{mt2015}.

\section{Calculation of thermodynamic quantities in two approaches}
Thus, we have two systems of equations: (\ref{1-6})--(\ref{1-8}) and
(\ref{1-9})--(\ref{1-13}). We will calculate from them the values of
$F$, $S$ and will compare the results. We consider only the regimes
of weak and infinitely strong couplings, for which the formulae for
$E(|p|)$ are available and the solutions for $F$ and $S$ can be
found in the $n$-approach (\ref{1-6})--(\ref{1-8})  easily.

1. Regime of weak coupling: $\gamma = c/n \ll 1$. For the
$n$-approach, we need to know the dispersion law of quasiparticles
\cite{mt2015} and the ground-state energy \cite{ll1963}. They are
given by the Bogolyubov's formulae for point particles:
\begin{equation}
E(p)= \sqrt{p^{4}+4cnp^{2}}, \quad E_{0}=Ncn(1-4\sqrt{\gamma}/(3\pi)).
     \label{1-14} \end{equation}
Let us substitute $E(p)$ in (\ref{1-6}). After the integration by
parts and a transformations, we get
\begin{equation}
F = E_{0} -\frac{\pi T^{2}L}{6v_{s}}I_{f}(a), \quad
a=\frac{v_{s}^{2}}{T}, \quad v_{s}=2\sqrt{cn},    \label{1-15}
\end{equation}
\begin{equation}
I_{f}(a) = \frac{6}{\pi^{2}
}\int\limits_{0}^{\infty}\frac{dx}{\sqrt{1+x^{2}/a^{2}}}\frac{x+2x^{3}/a^{2}}{e^{x\sqrt{1+x^{2}/a^{2}}}-1}.
     \label{1-16} \end{equation}
Here and below, $v_{s}$ is the velocity of sound.  Analogously,
formulae (\ref{1-8}), (\ref{1-6}), and (\ref{1-14}) yield
\begin{equation}
S = \frac{\pi T L}{3v_{s}}I_{s}(a), \quad I_{s}(a) =
\frac{6}{\pi^{2}}\int\limits_{0}^{\infty}
\frac{dx}{\sqrt{1+x^{2}/a^{2}}}\frac{x+3x^{3}/(2a^{2})}{e^{x\sqrt{1+x^{2}/a^{2}}}-1}.
     \label{1-17} \end{equation}
The values of $I_{f}(a)$ and $I_{s}(a)$ are presented in Fig. 1. For
$a\gg 1,$ we have $I_{f}\approx I_{s}\approx 1$.


\begin{figure}
\includegraphics[width=.6\textwidth]{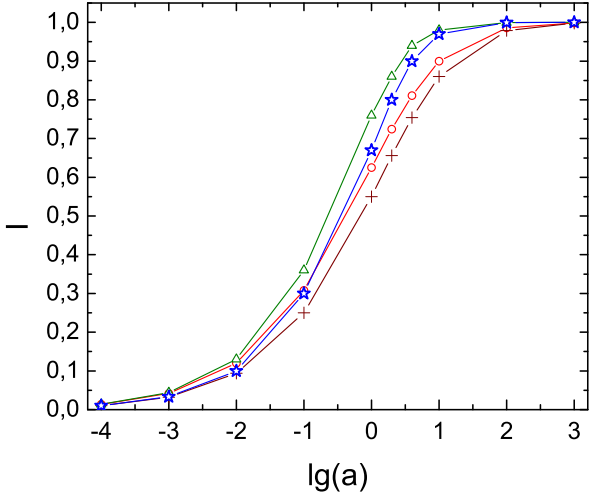}
\caption{[Color online] Functions $I_{f}(a)$
(${\sss\triangle\triangle\triangle}$),  $I_{s}(a)$
($\star\star\star$),   $I^{\infty}_{f}(a)$ ($\circ\circ\circ$), and
$I^{\infty}_{s}(a)$ ($+++$), see Eqs. (\ref{1-16}), (\ref{1-17}),
(\ref{1-25}), and (\ref{1-26}), correspondingly. In this and other
figures, $\lg{(x)}\equiv\log_{10}{x}$.}
          \label{fig:1}                  \end{figure}

Formulae (\ref{1-6})--(\ref{1-8}) hold if the number of
quasiparticles $N_{qp}$ is much less than the number of particles
$N$ \cite{mt2015}. The criterion is as follows: $N_{qp}\lsim 0.1N$.
The numerical  solution of Eqs. (\ref{1-4}) indicates that, for
$N_{qp}\gsim 0.1N,$ the interaction of quasiparticles changes
noticeably their energy. Therefore, formulae (\ref{1-5}) and
(\ref{1-6}) become not quite true. We now find $N_{qp}$. With regard
for (\ref{1-5}), the total internal energy of the system is
\begin{equation}
U=F+TS=F-T\frac{\partial F}{\partial
T}|_{N,V=const}=E_{0}+\sum\limits_{l\neq
0}\frac{E(l)}{e^{E(l)/T}-1}.
     \label{1-u} \end{equation}
On the other hand, since
\begin{equation}
U=E_{0}+\sum\limits_{l\neq 0}E(l)\bar{N}_{l},
     \label{1-u2} \end{equation}
we can find the mean number of Bose quasiparticles with energy
$E(l)$:
\begin{equation}
\bar{N}_{l}=(e^{E(l)/T}-1)^{-1}.
     \label{1-n} \end{equation}
From whence, the total number of quasiparticles is
\begin{equation}
N_{qp} =\sum\limits_{l \neq 0}\bar{N}_{l}=
\sum\limits_{p=-\infty}^{\infty}(e^{E(p)/T}-1)^{-1}|_{p\neq 0}=
2\sum\limits_{j=1}^{\infty}\left
(e^{\frac{v_{s}p_{1}j\sqrt{1+j^{2}p_{1}^{2}/v_{s}^{2}}}{T}}-1 \right
)^{-1} \approx \frac{2}{q_{1}}\ln{Q^{-1}},
     \label{1-18} \end{equation}
where $p=2\pi l/L$, $Q=q_{1}$ for $q_{1}\geq q_{2}$ and $Q=q_{2}$
for $q_{1}< q_{2}$, $q_{1}=v_{s}p_{1}/T$, $q_{2}=p_{1}/v_{s}$,
$p_{1}=2\pi/L$; it is assumed that $N, L \rightarrow \infty,
N/L=const$. The condition $N_{qp}\ll N$ requires
\begin{equation}
 \frac{T}{2n^{2}\sqrt{\gamma}}=\xi \ll \frac{\pi}{\ln{Q^{-1}}}.
     \label{1-19} \end{equation}
For large $L$ and $N,$ condition (\ref{1-19}) can be written as
$\frac{T}{2n^{2}\sqrt{\gamma}}\ll
\frac{\pi}{\ln{(\sqrt{\gamma}N)}}$. Condition (\ref{1-19}) indicates
that, for large $L$ and $N,$ the temperature $T$ should be low.
Relation (\ref{1-19}) implies that the value of $T$ decreases, if
$L$ increases. For $L=\infty,$ we have $T=0$. This is absurd result.
Why did such paradox appear? Formula (\ref{1-n}) follows from
(\ref{1-5}), and formula (\ref{1-5}) is derived from the canonical
Gibbs distribution. This distribution is obtained usually
\cite{huang,land5} from Liouville's theorem for an ensemble of
identical closed equilibrium systems and the division of each such
system into a small subsystem and the much larger thermostat. But if
our system is infinite and unbounded, it cannot be a part of the
``much larger'' system (even the imaginary one) of the same
dimension. Therefore, the Gibbs distribution is applicable only for
a finite system. In addition, the equilibrium arises due to the
interaction of parts of the system. Therefore, the equilibrium in a
system with infinite $N$ and $V$ is established in infinite time.
That is, the equilibrium in an infinite system makes no sense. In
the classical book by Gibbs \cite{gibbs}, the canonical distribution
was obtained from the condition of equilibrium in the system
[$\partial \rho(q_{1},\ldots,q_{N},p_{1},\ldots,p_{N},t)/\partial
t=0$] and Liouville's theorem for an ensemble of systems ($d \rho/d
t=0$). Gibbs noted that only the consideration of systems with a
finite partition function has meaning \cite{gibbs}. However, the
infinite systems with infinite number of degrees of freedom and with
a realistic interparticle interaction are characterized, presumably,
by the infinite partition function. In our opinion, the above
paradox arose due to the application of the Gibbs distribution to an
infinite system. We may apply the Gibbs distribution to a finite
system and then pass to the infinite system in formulae (to simplify
calculations, e.g.). By returning then to the finite system, we
expect to get a reasonable results. However, in this case, we may
obtain unphysical results for some properties at $N, V= \infty$. In
work \cite{mt2015}, we used the canonical Gibbs distribution for
finite $N$ and $L$; then, we set $N=\infty$, $L=\infty$ and obtained
formula (\ref{1-5}). But we assumed that this formula holds also for
finite $N$ and $L$.

\begin{figure}
\includegraphics[width=.6\textwidth]{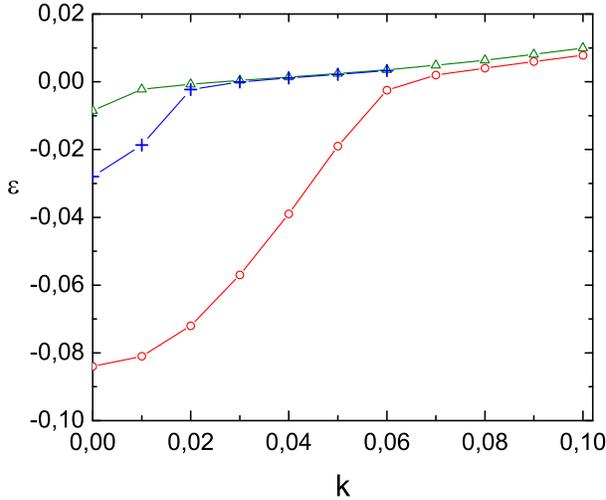}
\caption{[Color online] Function $\epsilon(k)$ as a solution of the
Yang--Yang equations (\ref{1-21}) and (\ref{1-10}) for $c=0$ (${\sss
\triangle\triangle\triangle}$), $c=0.0001$ ($+++$), and $c=0.001$
($\circ\circ\circ$). For $k>0.05,$ the curves approach one another
and the asymptote $\epsilon=k^{2}-\mu$. All points are determined
for $T=0.001$ and $n=1$. Each curve corresponds to the own value  of
$\mu$, which is determined from Eq. (\ref{1-10}). For free bosons,
$\epsilon(k)=T\ln{(e^{\frac{k^{2}-\mu}{T}}-1)}$ \cite{yangs1969} and
$\mu=-2.064\cdot 10^{-7}$. For $c=0.0001$ and $0.001,$ we found
$\mu\approx 2c$.}
          \label{fig:2}                  \end{figure}
In the traditional approach, it is necessary to solve the integral
equation (\ref{1-11}). By making in this equation changes
$k\rightarrow -k$ and (under the sign of integral)  $q\rightarrow
-q$, we get the same equation for the function $\epsilon(-k)$. This
means that
\begin{equation}
\epsilon(-k) = \epsilon(k).
     \label{1-20} \end{equation}
Therefore, instead of (\ref{1-11}), we can solve the equation
\begin{equation}
\epsilon(k) = -\mu +k^{2}-\frac{Tc}{\pi }\int\limits_{0}^{\infty}dq \ln{\left (1+e^{\frac{-\epsilon(q)}{T}}\right )}
\left (\frac{1}{c^{2}+(k-q)^{2}}+ \frac{1}{c^{2}+(k+q)^{2}}\right ),
     \label{1-21} \end{equation}
where $k\geq 0$. We tried several numerical procedures, but only the
method of iterations worked (one needs usually about $500$
iterations). The solutions for $\epsilon(k)$ at $c\ll 1 $  are shown
in Fig. 2.

Let us find the chemical potential $\mu$ for the known concentration
$n.$ We need to find $\epsilon(k)$ for $\mu$ and $\mu +\delta\mu$
with a small $\delta\mu$ and to substitute those $\epsilon(k)$ in
Eqs. (\ref{1-9}) and (\ref{1-10}): $n=\frac{P(\mu
+\delta\mu)-P(\mu)}{\delta\mu}|_{T=const}$. Analogously, we  obtain
the entropy from (\ref{1-9}) and (\ref{1-13}):
 $S=L\frac{P(T+\delta T)-P(T)}{\delta T}|_{\mu=const}$.
 The free energy can be found from (\ref{1-9}). Since
 formula (\ref{1-15}) is valid for low temperatures, the term
$\frac{\pi T^{2}L}{6v_{s}}I_{f}$ in (\ref{1-15}) is usually small as
compared with $E_{0}$. However, the entropy $S$ is determined namely
by this small term. Therefore, we now calculate the entropy, which
allow us to verify formula (\ref{1-15}) to within a small correction
$\sim T^{2}$.

In Figs. 3 and 4, we show the solutions for $S(c)$ and $S(T)$ found
in the traditional and new approaches. As is seen, both approaches
give the \textit{identical} solutions (a difference of $1$--$2\%$ is
connected with errors of the numerical method). For almost all
points in Figs. 3 and 4, we have $\xi\ll 1$, so that criterion
(\ref{1-19}) is satisfied (if $L$ is large, but not too large; $\xi$
is not small ($\xi\lsim 0.5$) near the point $T=0.1$ in Fig. 4). We
have $\mu \approx 2c$ for all points in Fig. 3 and $\mu\approx 0.02$
for all points in Fig. 4. We also derived the solutions $F(n)$ and
$S(n)$ for fixed $c= T =0.001$. In this case, for $n=0.5$--$2,$ we
have $\mu \approx n/500$, and the solutions for $S$ in the
traditional and new approaches coincide. The solutions for $F$
coincide as well.

In the $n$-approach, $\mu$ is zero, because the quasiparticles with
the Bose statistics can be freely created and annihilated. All
nonzero $\mu$ are related to the Yang--Yang approach, where the
system is described in the one-particle language.

%
%
%
 \begin{figure}
\centering\epsfig{figure=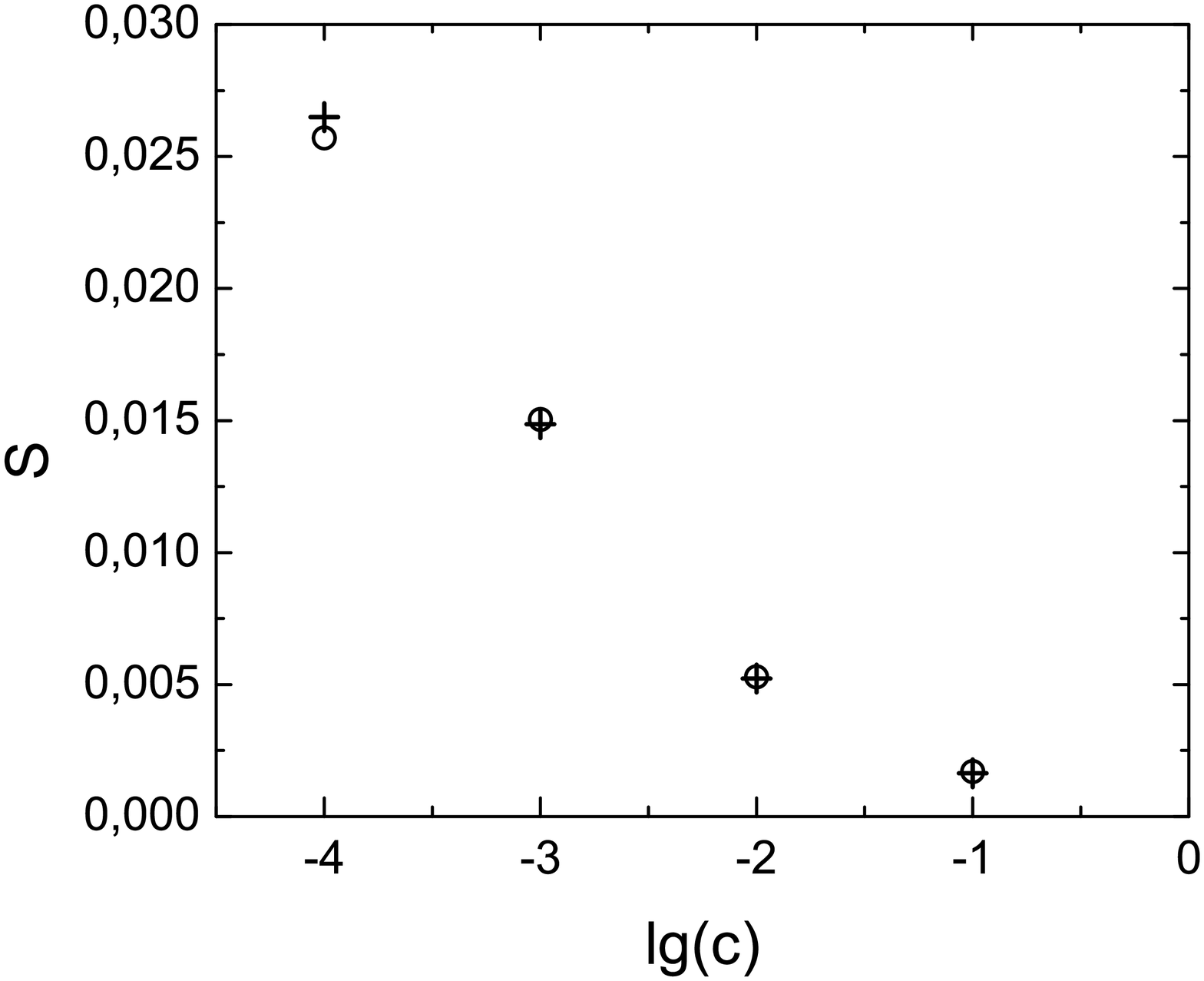,width=0.49\textwidth} \hfill
\centering\epsfig{figure=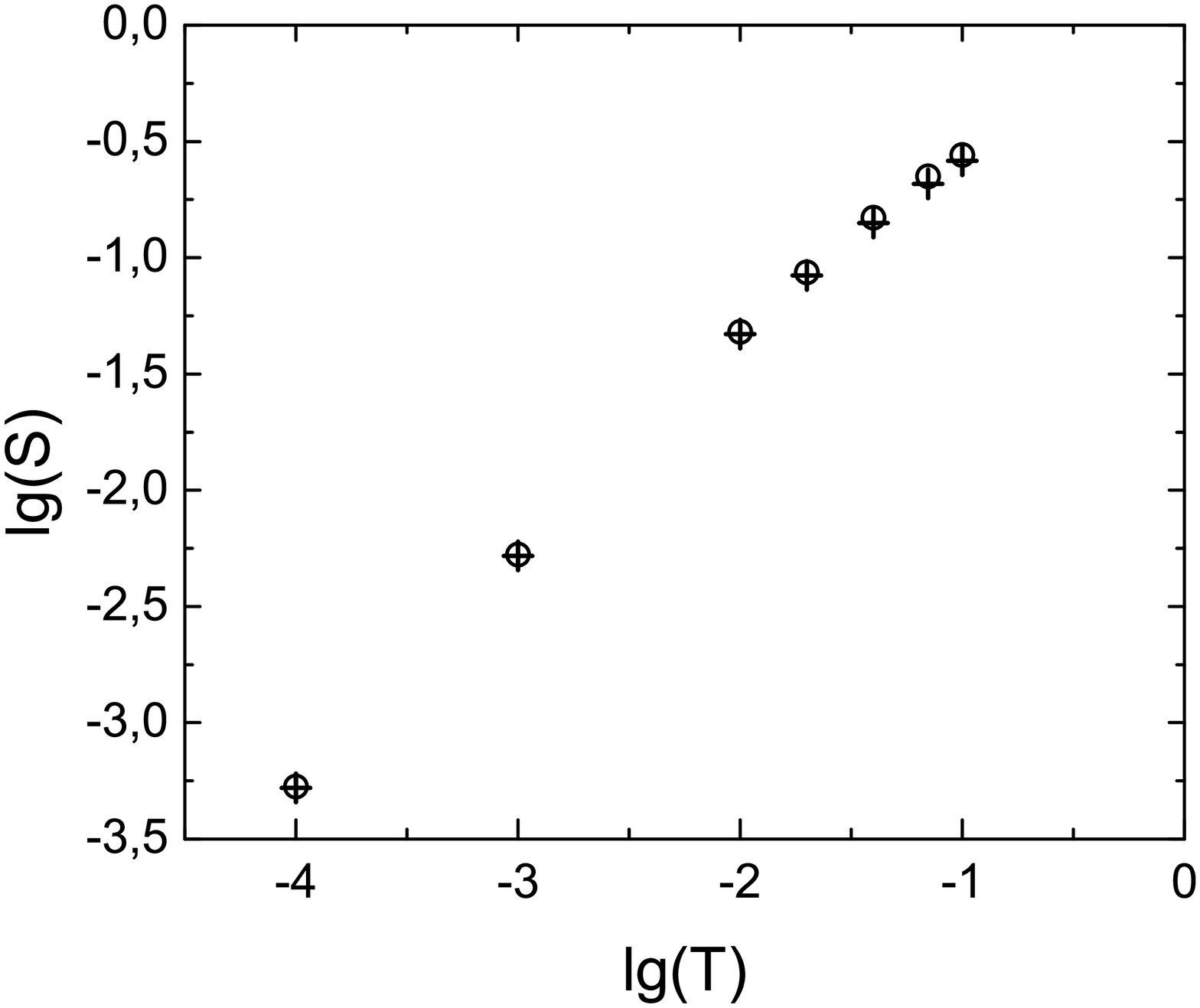,width=0.49\textwidth}
\\
\parbox[t]{0.47\textwidth}{
\caption{ Function $S(c)$ derived for the regime of weak coupling in
the traditional ($\circ\circ\circ$) and new ($+++$) approaches. For
all points, we have $n=1$ and $T=0.001$. The traditional approach
corresponds to Eqs. (\ref{1-20}), (\ref{1-21}), (\ref{1-10}), and
(\ref{1-13}), and the new one corresponds to formulae (\ref{1-17}).}
} \hfill
\parbox[t]{0.47\textwidth}{
\caption{ Function $S(T)$ obtained for the regime of weak coupling
($n=1$, $c=0.01$) in the traditional ($\circ\circ\circ$) and new
($+++$) approaches.} }
\end{figure}

For $c=0.0001$-$0.1$, we derived the  solution $\mu \approx 2c$ at
$n=1,$ $T=0.001$. This is of interest, because this requires $\mu
\approx 0$ for $c= 0$. However,  Eqs. (\ref{1-9})--(\ref{1-13})
yield for $c=0$ the equations for free bosons (they are given in
\cite{yangs1969}). In this case, $\mu$ is a solution of Eq.
(\ref{1-10}) and is \textit{negative} for all $n$ and $T.$ For
example, for $n=1$ and $T=0.001$ (parameters of Fig. 3), we get
$\mu\approx -2.064\cdot 10^{-7}< 0$. These results show that, most
likely, (1) $\mu>0$ for all $c>0$ and (2) at $c=0$, the value of
$\mu$ decreases \textit{by jump} to some $\mu <0$ (hence, $F$ and
$S$ vary also by jump at $c=0$). Though a smooth transition from a
positive $\mu(c=0.0001)$ to a negative $\mu(c=0),$ as $c$ decreases
from $0.0001$ down to $0,$ is also possible.  Fig. 2 shows the
solutions for $\epsilon(k)$ at $c=0; 0.0001,0.001$; these solutions
admit both possibilities. At both smooth and jump-like transition
$\mu(c>0.0001)\rightarrow \mu(c=0)$, the value of $\mu$ turns to
zero at some $c=c_{cr}\geq 0$. But we failed to find $c_{cr}$. To
clarify whether a jump exists, it is necessary to find a solution of
the Yang--Yang equations on the set of points of the domain
$0<c<0.0001$.

In the new approach, the situation is as follows.  For $\gamma \ll
1$, $c\rightarrow 0,$ and $n,T=const,$ we have $a=4cn/T\rightarrow
0$. The numerical analysis indicates that, at $a\rightarrow 0,$
$I_{f}\approx 1.4\sqrt{a}$, $I_{s}\approx 1.05\sqrt{a}$. Therefore,
formulae (\ref{1-15})--(\ref{1-17}) yield $F \approx E_{0} -1.4\pi
LT^{3/2}/6$, $S \approx 1.05\pi LT^{1/2}/3$. Since $E_{0}(c=0)=0$,
we have $F\approx-2TS/3$. For free bosons, $F=-TS$ (this relation
can be obtained from formulae in \cite{yangs1969}). The difference
of the factors $2/3$ and $1$ is related to the jump of $F$ and (or)
$S$ or to the fact that, at $c\rightarrow 0,$ condition (\ref{1-19})
is violated, and formulae (\ref{1-15})--(\ref{1-17}) become not
quite proper.

The jump of $F$ and (or) $S$ at $c=0$ is possible due to the
transition to the thermodynamic limit. Indeed, for a finite system,
the minimal $|p|$ in the Bogolyubov formula $E(p)=
\sqrt{p^{4}+4cnp^{2}}$ is equal to $|p|= 2\pi/L$. If we pass to the
thermodynamic limit for arbitrarily small $c,$ then $L$ can be taken
so large that, at smallest $|p|$, the relation $p^{4}\ll 4cnp^{2}$
will hold. Therefore, the dispersion law will be linear in $p$,
which yields formulae (\ref{1-15})--(\ref{1-17}). At fixed finite
$L,$ the value of $c$ could be taken so small that the dispersion
law would be quadratic in $p$ at small $p,$ like $E(p)$ for free
particles. Therefore, the thermodynamic solutions for  small $c$
would undoubtedly pass continuously to the solutions for $c=0$.
However, for an infinite system, the point $c=0$ is singular for the
dispersion law and, therefore, can be singular for the thermodynamic
quantities as well. We recall that the above-discussed paradoxical
conclusion that the condition $N_{qp}\ll N$ requires $T=0$ is also
related to the transition to the thermodynamic limit.

It is worth noting that, for a 1D system of point bosons, an analog
of the phase transition at the point $\mu=0$  was found
\cite{batchelor2011,guan2015} for the regime $\gamma\gg 1$. As far
as we see, the analyticity of the thermodynamic functions in $\mu$
and $T$ is conserved in a vicinity of the point $\mu=0$, which
corresponds to the proof \cite{yangs1969}. But it is unclear whether
this peculiarity reveals in the $n$-approach (for which $\mu$ can be
only zero, because the physics is determined by  Bose
quasiparticles) and, if yes, what is the physical meaning of this
peculiarity?

\begin{figure}
\includegraphics[width=.6\textwidth]{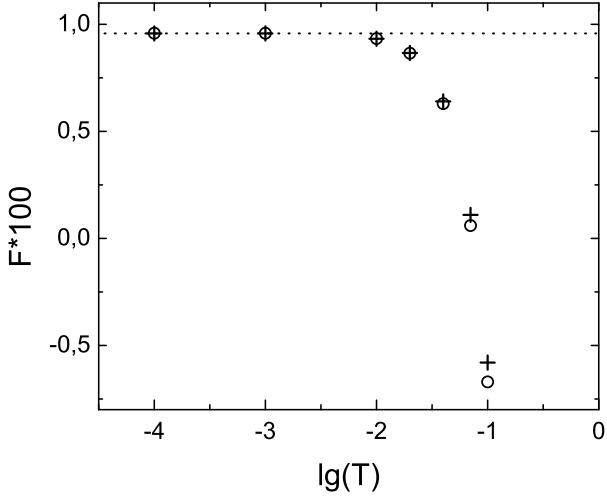}
\caption{Function $F(T)$ derived for the regime of weak coupling
($n=1$, $c=0.01$) in the traditional ($\circ\circ\circ$) and new
($+++$) approaches. The traditional approach corresponds to Eqs.
(\ref{1-20}), (\ref{1-21}), (\ref{1-10}), (\ref{1-9}), and the new
one --- to formulae (\ref{1-15}), (\ref{1-16}).  The value of $F$ is
increased by 100 times. The dotted line shows the asymptote
$F(T\rightarrow 0)=E_{0}=0.00958$.}
          \label{fig:5}                  \end{figure}


In Fig. 5, we present the curves $F(T)$ found in the new and
traditional approaches. These curves practically coincide. In
particular, the curve $F(T)$ of the Yang--Yang method approaches at
$T\rightarrow 0$ the asymptote $F=E_{0}$ corresponding to formula
(\ref{1-15}) of the $n$-approach. With the increase in $T$, the
results of the traditional and new solutions become somewhat
different. This is related to the fact that  $\xi$ becomes large
(e.g., for $T=0.1,$ we have $\xi\approx 0.5$), and therefore,
condition (\ref{1-19}) is broken. For the points with $T\lsim 0.02$
we have $\xi\lsim 0.1$.

2. Regime of infinitely strong coupling: $c =+\infty$, $\gamma
=+\infty$. Under the description in the language of atoms, it is the
Fermi-like regime \cite{yangs1969} (though the wave function has the
Bose symmetry). But, under the description in the language of
$n$-quasiparticles \cite{mt2015}, we have the purely bosonic regime.

First, we consider the $n$-approach \cite{mt2015}. The dispersion
law of quasiparticles and the ground-state energy are determined by
the Girardeau's formulae \cite{girardeau1960}:
\begin{equation}
E^{\infty}(p)= p^{2}+2\pi n|p|, \quad
E^{\infty}_{0}=Nn^{2}\pi^{2}/3.
     \label{1-23} \end{equation}
Substitute $E^{\infty}(p)$ in (\ref{1-6}) and make some
transformations, then we obtain:
\begin{equation}
F = E^{\infty}_{0} -\frac{\pi
T^{2}L}{6v^{\infty}_{s}}I^{\infty}_{f}(a^{\infty}), \quad
a^{\infty}=\frac{(v^{\infty}_{s})^{2}}{T}, \quad v^{\infty}_{s}=2\pi
n,
     \label{1-24} \end{equation}
\begin{equation}
I^{\infty}_{f}(a^{\infty}) = \frac{6}{\pi^{2}
}\int\limits_{0}^{\infty}
dx\frac{x+2x^{2}/a^{\infty}}{e^{x(1+x/a^{\infty})}-1}.
     \label{1-25} \end{equation}
Formulae (\ref{1-8}), (\ref{1-6}), and (\ref{1-23})  yield
\begin{equation}
S = \frac{\pi T L}{3v^{\infty}_{s}}I^{\infty}_{s}(a^{\infty}), \quad
I^{\infty}_{s}(a^{\infty}) =
\frac{6}{\pi^{2}}\int\limits_{0}^{\infty}
dx\frac{x+3x^{2}/(2a^{\infty})}{e^{x(1+x/a^{\infty})}-1}.
     \label{1-26} \end{equation}
Formulae for $F$ and $S$ are \textit{the same} as those for the
regime of weak coupling. The difference consists only in the changes
$E_{0}\rightarrow E^{\infty}_{0}$, $v_{s}\rightarrow
v^{\infty}_{s}$, $I_{f}(a)\rightarrow I_{f}^{\infty}(a^{\infty})$,
$I_{s}(a)\rightarrow I_{s}^{\infty}(a^{\infty})$. The values of
$I_{f}^{\infty}(a^{\infty})$ and $I_{s}^{\infty}(a^{\infty})$  are
shown in Fig. 1. We have $I^{\infty}_{f}\approx
I^{\infty}_{s}\approx 1$ for $a^{\infty}\gg 1$ and
$I^{\infty}_{f}\approx 1.4\sqrt{a^{\infty}}$ ,
$I^{\infty}_{s}\approx 1.04\sqrt{a^{\infty}}$ for $a^{\infty}\ll 1$.

Formulae (\ref{1-24}), (\ref{1-26}) with
$I^{\infty}_{s}=I^{\infty}_{f}=1$  were obtained previously
\cite{taka2007,batchelor2011} for the regime $T\rightarrow 0$,
$\gamma \rightarrow \infty$  in a more complicated way from the
Yang--Yang equations \cite{yangs1969} (in works
\cite{taka2007,batchelor2011}, there is the small slip in the
formula for $v_{c}$ (our $v^{\infty}_{s}$): It should be
$v^{\infty}_{s}=(\frac{1}{m}\frac{\partial P}{\partial
n})^{1/2}=\frac{\hbar n \pi}{m}=2\pi n$, then the formula for $F(T)$
in \cite{taka2007,batchelor2011} coincides with (\ref{1-24}) with
$I^{\infty}_{f}=1$). Still before, the formula for $F$, close to
(\ref{1-24}), was obtained by the field-theoretic method
\cite{affleck1986}.

Let us find out the consequences of the condition $N_{qp}\ll N$. At
$N, L\rightarrow \infty$, we obtain
\begin{equation}
N_{qp} =
\sum\limits_{p=-\infty}^{\infty}(e^{E^{\infty}(p)/T}-1)^{-1}|_{p\neq
0}= 2\sum\limits_{j=1}^{\infty}\left (e^{q_{1}j(1+q_{2}j)}-1 \right
)^{-1} \approx\frac{2}{q_{1}}\ln{Q^{-1}},
     \label{1-27} \end{equation}
where $Q=max(q_{1},q_{2})$, $q_{1}=v^{\infty}_{s}p_{1}/T$,
$q_{2}=p_{1}/v^{\infty}_{s}$, $p_{1}=2\pi/L$. Relation $N_{qp}\ll N$
yields
\begin{equation}
T\ll \frac{(v^{\infty}_{s})^{2}}{2\ln{Q^{-1}}}.
     \label{1-28} \end{equation}
For not too large $L,$ we have $\ln{Q^{-1}}\sim 10$, and
(\ref{1-28}) is approximately reduced to $T \ll n^{2}$.

Let us turn to the Yang--Yang approach.  For $c=\infty,$ Eq.
(\ref{1-11}) has the solution $\epsilon(k) = -\mu +k^{2}$. Equations
(\ref{1-9})--(\ref{1-13}) yield the formulae
\begin{equation}
\quad P = \frac{2T\sqrt{T}}{\pi }\int\limits_{0}^{\infty}dq
\frac{q^{2}}{e^{q^{2}-\eta}+1},
     \label{1-29} \end{equation}
\begin{equation}
n=\frac{\partial P}{\partial
\mu}|_{T=const}=\frac{\sqrt{T}}{\pi}\int\limits_{0}^{\infty}\frac{dq}{e^{q^{2}-\eta}+1},
     \label{1-30} \end{equation}
\begin{equation}
S=L\frac{\partial P}{\partial T}|_{\mu=const}=\frac{N\sqrt{T}}{\pi n
}\int\limits_{0}^{\infty}dq  \frac{3q^{2}-\eta}{e^{q^{2}-\eta}+1},
     \label{1-31} \end{equation}
where $\eta=\mu/T$.  In this case,
$S=\frac{3LP}{2T}-\frac{N\mu}{T}$. Therefore, we have
$PL=\frac{2ST}{3}+\frac{2N\mu}{3}$, which yields
\begin{equation}
F = N\mu -PL = \frac{NT}{3 }\left (\eta-\frac{2S}{N}\right ).
     \label{1-32} \end{equation}
Let the concentration $n$ be known. Then, we numerically derive
$\eta$ from Eq. (\ref{1-30}) and $S$, $F$ from (\ref{1-31}),
(\ref{1-32}). For $T\rightarrow 0$ and $\mu>0$ in (\ref{1-30}), we
have $e^{q^{2}-\eta}\rightarrow 0$ for all $q^{2}<\eta$. Therefore,
$\eta(T\rightarrow 0)=\pi^{2} n^{2}/T+\varphi (T/n^{2})$. The
numerical analysis indicates that $\varphi (\frac{T}{n^{2}}
\rightarrow 0)\approx 0.1\frac{T}{n^{2}}$. At $T\rightarrow 0$ and
$\mu\leq 0,$ Eq. (\ref{1-30}) yields $n \rightarrow 0$.

In Fig. 6, we show the solution $S(T)$ found numerically from Eqs.
(\ref{1-30}), (\ref{1-31})  as compared with solution (\ref{1-26})
in the $n$-approach. It is seen that both solutions coincide with
good accuracy for $T< 1$ and are slightly different for $T\gsim 1$.
The last is because the ratio $N_{qp}/N$ becomes large (of the order
of 1 for $N=10^{4}, n=1$) for $T\gsim 1.$ Therefore, the
approximation of free quasiparticles \cite{mt2015} becomes improper.
For $T\leq 0.01,$ the difference of solutions (\ref{1-26}) and
(\ref{1-31}) for $S(T)$ is at most $0.1\%$. Solutions for $F(T),$
(\ref{1-24}) and (\ref{1-32}), coincide for small $T$ with even
higher accuracy (we do not portray them).

\begin{figure}
\includegraphics[width=.6\textwidth]{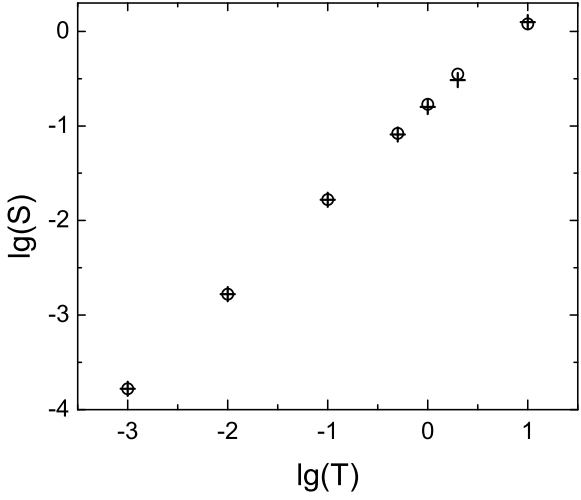}
\caption{Function $S(T)$ for the regime of strong coupling
($c=+\infty$, $n=1$) obtained in the traditional ($\circ\circ\circ$)
and new ($+++$) approaches. The traditional approach corresponds to
Eqs. (\ref{1-30}) and (\ref{1-31}), and the new one --- to formulae
(\ref{1-26}).}
          \label{fig:6}                  \end{figure}


 For the regime of strong coupling, the entropy of the
quasi-1D Bose gas was measured \cite{ott2013} and approximately
agrees with the solution of the Yang--Yang equations.

For small $T,$ the thermodynamics is defined by the sound part of
the dispersion law. Therefore, formulae (\ref{1-15}) and
(\ref{1-17}) (with $I_{s}=I_{f}=1$ and the value of $v_{s}$
corresponding to $\gamma$ under consideration) should be valid for
\textit{any} coupling constant $\gamma$.

A 1D system of fermions with the holon and spinon excitations was
considered in \cite{batchelor2012}. For small $T$ and any coupling
constant, the following formula was obtained \cite{batchelor2012}:
\begin{equation}
F = E_{0} -\frac{\pi T^{2}L}{6}\left
(\frac{1}{v^{(1)}_{s}}+\frac{1}{v^{(2)}_{s}}\right ).
     \label{1-33} \end{equation}
It is a natural generalization of formula (\ref{1-15}) to the case
of a system with two types of quasiparticles (the holon dispersion
curve is gapped, the spinon one is gapless; in both cases,
$v^{(i)}_{s}=\frac{\partial E^{(i)}}{\partial
k}|_{E^{(i)}\rightarrow 0}$). Therefore, we  assume that the
$n$-approach is also applicable to other integrable systems.

Thus, for $\gamma \ll 1$ and $\gamma=+\infty,$ the solutions in the
new approach \cite{mt2015} coincide with the corresponding solutions
in the Yang--Yang approach \cite{yangs1969}. We have no doubts that
such coincidence holds also for the intermediate values of $\gamma.$

  \section{Discussion of the results and experiments}
The thermodynamics of a 1D system of point bosons can be constructed
within the traditional Yang--Yang method and the new $n$-method
\cite{mt2015}. These approaches are equivalent. In the previous
section, we have seen that they lead to the same results for $F$ and
$S$. However, the new approach seems to be somewhat more physical,
in the following sense. At small temperatures, any excited state of
a gas is most simply described as a set of quasiparticles. The
$n$-approach is constructed namely in the language of quasiparticles
and, therefore, leads to simpler equations. In this case, the
quasiparticles are characterized by the same statistics, as the
quasiparticles in a system of nonpoint bosons. As for the Yang--Yang
approach, it applies the language of individual atoms, and the
formulae do not correspond to a definite statistics, generally
speaking. However, the Yang--Yang approach is more universal, since
it allows one to find the thermodynamic quantities at any
temperature (the $n$-approach works only at small $T$).

In the recent years, some interesting experiments with a quasi-1D
Bose gas in a trap were carried out \cite{nagerl2015,inguscio2015}.
The application of the Bragg spectroscopy allowed one to get the
detailed experimental data on a dynamical structural factor. In
particular, those data give information about the dispersion law for
quasiparticles. The authors of works \cite{nagerl2015,inguscio2015}
made conclusion about the essential contribution of the hole-like
and particle-like excitations to the scattering, because at $\gamma
> 3$ the scattering peak  is placed \textit{between} the energy of a
particle-like excitation and the energy of a hole-like excitation
\cite{nagerl2015}. According to the conclusion \cite{inguscio2015},
the inhomogeneity of the system affects slightly the peak width, the
broadening has a non-temperature nature and is related to the
interaction (of separate atoms, apparently).

Our impression from the results \cite{nagerl2015,inguscio2015} is
the following. The vibrations or rotations of the cloud as a whole
have an insignificant influence on the Bragg spectroscopy; the
inhomogeneity of a gas affects slightly the peak width
\cite{inguscio2015}. Therefore, the physics of the system should be
defined by quasiparticles, like for a uniform quantum liquid, such
as He II. That is, the broadening of the experimental peak is
related, in our opinion, to the usual temperature mechanism
(interaction of quasiparticles). It is  interesting that the
experimental peak deviates from the energy of particle-like
quasiparticles (see Fig. 2 in \cite{nagerl2015}), if $\gamma$
increases. We recall that the $n$-approach is equivalent to the
Lieb's and Yang--Yang approaches. Moreover, a Lieb's quasiparticles
(of both types) can be presented as one or several
$n$-quasiparticles. The energy of Lieb's particle-like excitation
coincides with the energy of $n$-quasiparticle \cite{mt2015} (at the
same momentum, of course). But the $n$-approach involves only
$n$-quasiparticles. Therefore, it is strange that the experimental
peak deviates from the energy of this quasiparticle. This can be
related to not quite accurate determination of some parameters of
the system or to the too simplified description of the system. The
more radical possibility consists in that the solutions for point
bosons do not coincide with the solutions for real nonpoint bosons.
However, the observation of only one peak \cite{nagerl2015} agrees
with the $n$-picture, because the last contains one type of
excitations. If the system would have two independent types of
excitations, then we would observe two peaks.

The picture with ``holes'' and ``particles'' took deep root, but our
approach is simpler and presumably more physical. Therefore, it is
worth attempting to interpret the experimental data in the language
of this approach.

It is also noteworthy that a hole-like excitation, which presents
several co-directed phonons (according to the $n$-approach), is
related to solitons
\cite{hole-soliton1,hole-soliton2,sato2012,sato2016}.

The future problem is the construction of the thermodynamics for a
1D system of finite size. For the infinite system, the thermodynamic
quantities depend on $T$ analytically \cite{yangs1969}. Does this
analyticity conserve also for a finite system? For a 1D system of
point bosons, the number of quasiparticles does not exceed the
number of atoms. Therefore, for a finite system, we must take
$\eta_{l}=0,1,2,\ldots, N$ and $\sum\limits_{l}\eta_{l}\leq N$ in
the partition function (see Eq. (65) in \cite{mt2015}). Such sum is
easily calculated only for $N=\infty$ \cite{mt2015}.

 \section{Summary}
Since the first works \cite{girardeau1960,ll1963,lieb1963,yangs1969}
till now, the one-dimensional system of spinless point bosons is
described in the language of fermions. The fermionicity is
manifested in the properties of Eqs. (\ref{1-3}), presence of
hole-like excitations, and Fermi-like structure of the thermodynamic
equations \cite{yangs1969}. In the present work and in
\cite{mt2015}, we have shown that the point bosons can be described
in the purely bosonic language, by using Eqs. (\ref{1-4}) instead of
equivalent ones (\ref{1-3}). In our approach, we have only one type
of quasiparticles. At a weak coupling, they are Bogolyubov
quasiparticles. One succeeded in constructing the thermodynamics for
the infinite system in the language of quasiparticles \cite{mt2015}.
In this case, the method is essentially different from the
Yang--Yang method \cite{yangs1969} and give the ordinary formulae
for an ensemble of noninteracting Bose quasiparticles, like for a
system of nonpoint bosons (e.g., for He II). It is not too strange,
because the point bosons are the limiting case of nonpoint ones. In
the present article, we have shown that the solutions for the
thermodynamic quantities in the new \cite{mt2015} and traditional
\cite{yangs1969} approaches \textit{coincide}. Thus, a 1D system of
spinless point bosons can be described in both bosonic and fermionic
languages. This is of interest, since only the bosonic language is
developed for nonpoint bosons. Apparently, the approach
\cite{mt2015} can be applied also to other integrable systems. In
particular, some Fermi systems without a pairing can probably be
described in a bosonic language.

Moreover, we have found the evidence of a possible jump of the
thermodynamic quantities of infinite system at the coupling constant
$c=0$.

     \renewcommand\refname{}


\begin{thebibliography}{200}
\bibitem {lieb1963}  E.H. Lieb, Phys. Rev. \textbf{130}, 1616 (1963).
\bibitem {bog1947} N.N. Bogoliubov, J. Phys. USSR \textbf{11}, 23 (1947).
\bibitem {bz1956}  N.N. Bogoliubov and D.N. Zubarev, Sov. Phys. JETP \textbf{1}, 83 (1956).
\bibitem {fc} R.P. Feynman and M. Cohen, Phys. Rev. \textbf{102}, 1189 (1956).
\bibitem {brueckner}  K. Brueckner, {\it Theory of Nuclear Structure}
        (Methuen, London, 1959).
\bibitem {rea}  L. Reatto,  J.~Low Temp. Phys., \textbf{87},  375 (1992).
\bibitem {mtreview}  M.D. Tomchenko,  arXiv:0904.4434.
\bibitem {yangs1969}  C.N. Yang, C.P. Yang, J. Math. Phys. (N.Y.) \textbf{10}, 1115 (1969).
 \bibitem {huang} K. Huang, {\it Statistical Mechanics}
           (Wiley, New York, 1963), Chapters 4, 8, 9, 18.
\bibitem {mtcond2016}  M. Tomchenko,  J. Low Temp. Phys. \textbf{182}, 170 (2016).
\bibitem {gaudinm}  M. Gaudin, {\it The Bethe Wavefunction} (Cambridge Univ. Press, Cambridge, 2014).
\bibitem {takahashi} M. Takahashi, {\it Thermodynamics of One-Dimensional Solvable Models}
           (Cambridge Univ. Press, Cambridge, 1999).
\bibitem {essler2005}  F.H.L. Essler, H. Frahm, F. G\"{o}hmann, A.
     Kl\"{u}mper, and V.E. Korepin {\it The One-Dimensional Habbard Model}
    (Cambridge Univ. Press, Cambridge, 2005).
\bibitem {guan2013}  X.-W. Guan, M.T. Batchelor, C. Lee, Rev. Mod. Phys. \textbf{85}, 1633 (2013).
\bibitem {mt2015}  M. Tomchenko,  J. Phys. A: Math. Theor. \textbf{48}, 365003 (2015).
\bibitem {ying2001}  S.J. Gu, Y.Q. Li, Z.J. Ying,
         J. Phys. A: Math. Gen. \textbf{34}, 8995 (2001).
\bibitem {taka2007} H.-W. Guan, M.T. Batchelor, and M. Takahashi, Phys. Rev. A \textbf{76},  043617 (2007).
\bibitem {batchelor2011}   X.W. Guan, M.T. Batchelor, J. Phys. A: Math. Theor. \textbf{44}, 102001 (2011).
\bibitem {guan2015}  Y.-Z. Jiang, Y.-Y. Chen, and X.-W. Guan, Chin. Phys. B \textbf{24}, 050311 (2015).
\bibitem {ll1963}  E.H. Lieb, W. Liniger, Phys. Rev. \textbf{130}, 1605 (1963).
\bibitem {gaudin1971}  M. Gaudin, Phys. Rev. A \textbf{4}, 386 (1971).
\bibitem {land5} L.D.~Landau and E.M.~Lifshitz, {\it Statistical Physics},  Part 1
        (Pergamon Press, Oxford, 1980), Chapters 2, 3, 5, 6.
\bibitem {mt2016edinst}  M. Tomchenko, J. Phys. A: Math. Theor. \textbf{50}, 055203 (2017).
 \bibitem {gibbs} J.W.  Gibbs, {\it Elementary Principles in Statistical Mechanics}
           (Scribner's sons, New York, 1902), Chapters I, IV.
\bibitem {girardeau1960}  M. Girardeau, J. Math. Phys. (N.Y.) \textbf{1}, 516 (1960).
\bibitem {affleck1986}  I. Affleck, Phys. Rev. Lett. \textbf{56}, 746 (1986).
\bibitem {ott2013}  A. Vogler, R. Labouvie, F. Stubenrauch, G. Barontini, V. Guarrera,
                   and H. Ott, Phys. Rev. A \textbf{88}, 031603(R) (2013).
\bibitem {batchelor2012}  J.Y. Lee, X.W. Guan, K. Sakai, and M.T. Batchelor, Phys. Rev. B \textbf{85}, 085414 (2012).
\bibitem {nagerl2015}  F. Meinert, M. Panfil, M.J. Mark, K. Lauber, J.-S. Caux,
                   and H.-C. N\"{a}gerl, Phys. Rev. Lett. \textbf{115}, 085301 (2015).
\bibitem {inguscio2015}  N. Fabbri, M. Panfil, D. Clement, L. Fallani, M. Inguscio, C. Fort, and  J.-S. Caux,
                   Phys. Rev. A \textbf{91}, 043617 (2015).
\bibitem {hole-soliton1}  M. Ishikawa, H. Takayama, J. Phys. Soc. Jpn. \textbf{49}, 1242 (1980).
\bibitem {hole-soliton2}  T. Tsuzuki, J. Low Temp. Phys. \textbf{4}, 441 (1981).
\bibitem {sato2012}  J. Sato, R. Kanamoto, E. Kaminishi, and T. Deguchi,   arXiv:1204.3960.
\bibitem {sato2016}  J. Sato, R. Kanamoto, E. Kaminishi, and T. Deguchi, New J. Phys.  \textbf{18}, 075008 (2016).

\end{thebibliography}
       \end{document}